\documentclass[doublespace,fleqn,10pt]{wlscirep}
\usepackage[utf8]{inputenc}
\usepackage{graphicx}
\usepackage{mathtools}
\usepackage{xcolor}
\usepackage{amsmath}
\usepackage{amssymb}
\usepackage{xfrac}
\usepackage{soul}

\newcommand\rst{\bgroup\markoverwith{\textcolor{red}{\rule[0.5ex]{2pt}{0.4pt}}}\ULon}

\begin{document}

\title{Quantized Polarization Redefines Polar Interfaces}
\author[1]{Hongsheng Pang}
\author[1,2,3*]{Lixin He}

\affil[1]{Key Laboratory of Quantum Information, University of Science and
Technology of China, Hefei, 230026, People's Republic of China}
\affil[2]{Institute of Artificial Intelligence, Hefei Comprehensive National Science Center,
		Hefei, Anhui, 230026, People's Republic of China}
\affil[3]{Hefei National Laboratory, University of Science and
		Technology of China, Hefei, Anhui,  230088, People's Republic of China}

\affil[*]{helx@ustc.edu.cn}

\begin{abstract}	
In crystalline solids, the electronic polarization follows the \emph{generalized Neumann’s principle}, under which all crystallographic point groups can, in principle, support ferroelectric polarization. However, in high-symmetry structures, polarization is constrained by symmetry operations and becomes quantized into discrete values. We demonstrate that this quantized polarization (QP) is not a mathematical artifact but a \emph{symmetry-protected invariant} that encodes intrinsic information about a material’s symmetry and electronic structure. Because of its discrete and non-continuous nature, when two materials with different QPs form an interface, their bulk polarization states cannot be connected adiabatically, compelling the system to develop pronounced interfacial responses—such as metallic states, bound charges, or strong lattice distortions. This theoretical framework provides a unified reinterpretation of classical systems such as the LaAlO$_3$/SrTiO$_3$ interface, revealing it as a prototypical case of QP mismatch. By establishing QP as a fundamental bulk invariant, our work uncovers a universal mechanism governing interfacial electronic phenomena and opens new pathways for the design of functional quantum materials through engineered polarization mismatch.
\end{abstract}

\maketitle

\section*{Introduction}
Ferroelectricity, defined by the presence of switchable polarization, has been at the heart of condensed matter physics for nearly a century\cite{cochran1960,portengen1996,lines2001}.
Its discovery not only established a cornerstone in solid-state science but also enabled applications ranging from non-volatile memories to sensors and actuators~\cite{rabe2007physics}. The theoretical understanding of ferroelectricity has long been grounded in Neumann’s principle, which dictates that macroscopic properties must conform to the symmetry of the crystal’s point group\cite{neumann1885vorlesungen}. Within this framework, only ten of the thirty-two crystallographic point groups, the so-called polar groups, are compatible with a nonzero polarization, providing the conventional classification of ferroelectrics.

Recently, however, a qualitatively new class of ferroelectrics has been uncovered: \emph{fractional quantum ferroelectricity} (FQFE)~\cite{Ji2024, Yu2025,pang2025}. In these materials, polarization assumes quantized fractional multiples of the polarization quantum. This discovery significantly broadens the scope of ferroelectrics. While FQFE may initially seem to contradict the conventional Neumann’s principle, our recent work~\cite{pang2025} established that this paradox naturally dissolves within a \emph{generalized Neumann’s principle}, which incorporates translational symmetry and the quantum of polarization. Under this framework, all thirty-two crystallographic point groups, including centrosymmetric ones, can host well-defined polarization states~\cite{vanderbilt1993electric,vanderbilt2018berry}, where the fractional quanta of polarization arise due to the symmetry restriction~\cite{pang2025}.  Strikingly, even a cubic crystal can exhibit half-quantum polarization.
Yet, since such quantized polarizations (QPs) are generally regarded as non-switchable in many materials, their profound physical significance has largely been overlooked, with QPs treated merely as reference states in high-symmetry crystals.

\begin{figure}[tbp]
			\centering
			\includegraphics[width=0.4\linewidth]{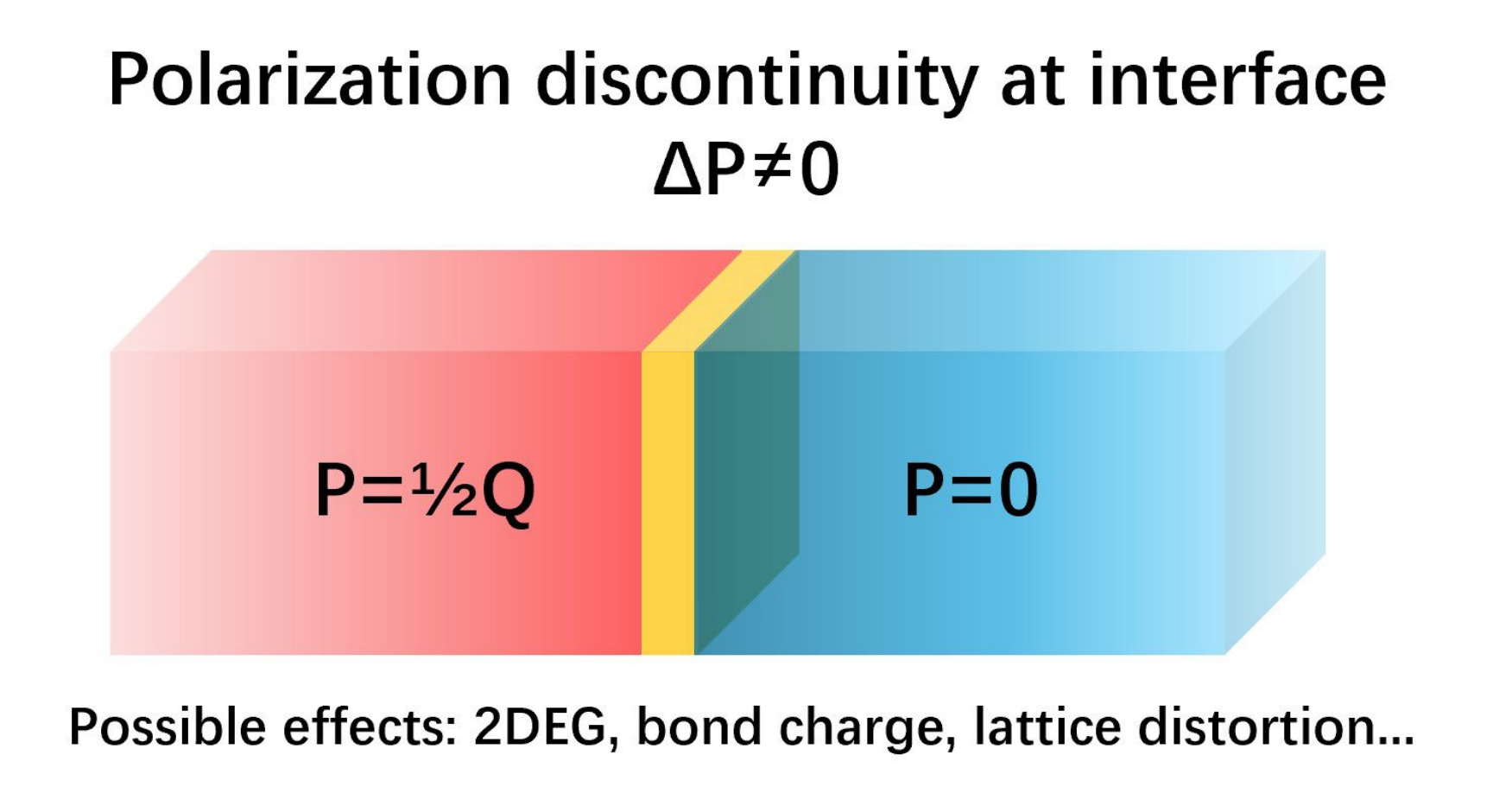}
			\caption{\textbf{QP mismatch drives emergent interfacial phenomena.} Schematic showing that an interface between materials with distinct QPs cannot be connected adiabatically, leading to polarization discontinuity and emergent interfacial phenomena such as two-dimensional electron gas (2DEG) or large lattice distortion etc.
			}
			\label{fig:interface}
		\end{figure}

Here we demonstrate that the QP is not a formal artifact, but a robust, symmetry-protected invariant of the crystal that encodes intrinsic information about its symmetry and electronic structure. As a quantized and discrete invariant, it cannot change continuously across space. This immutability has profound implications for material interfaces. When two materials with different QPs are brought together, their bulk polarization states cannot be connected through any adiabatic path, inevitably giving rise to induced interfacial phenomena, such as metallic interface states, bound charges, or pronounced lattice distortions, as illustrated in Fig.~\ref{fig:interface}.
Consequently, QP provides a symmetry-protected classification of materials that governs interfacial physics, analogous to the bulk–boundary correspondence in topological matter~\cite{chern,xiao2010berry}.
Furthermore, unlike conventional topological indices, discontinuities in QP typically introduce large electrostatic energy penalties, leading to much richer and more complex interfacial phenomena than those found in ordinary topological systems. In contrast, interfaces between materials with identical QPs can remain insulating and free from such anomalies.

Interestingly, the well-known metallic interfacial states at the LaAlO$_3$/SrTiO$_3$ interface~\cite{LAO-STO-1,LAO-STO-2,LAO-STO-3} can also be described within this theoretical framework, revealing that it represents a prototypical case of mismatched QPs. This establishes a general principle applicable to a broader class of materials, predicting analogous interfacial phenomena well beyond canonical oxide heterostructures. By elevating QP to a fundamental invariant, this work not only deepens the conceptual foundation of ferroelectricity but also opens new avenues for materials design, where engineered QP discontinuities and interface-induced metallic states may serve as the functional basis for next-generation electronic devices.

\section*{Results and Discussion}

		\begin{figure}[tbp]
			\centering
			\includegraphics[width=0.45\linewidth,keepaspectratio]{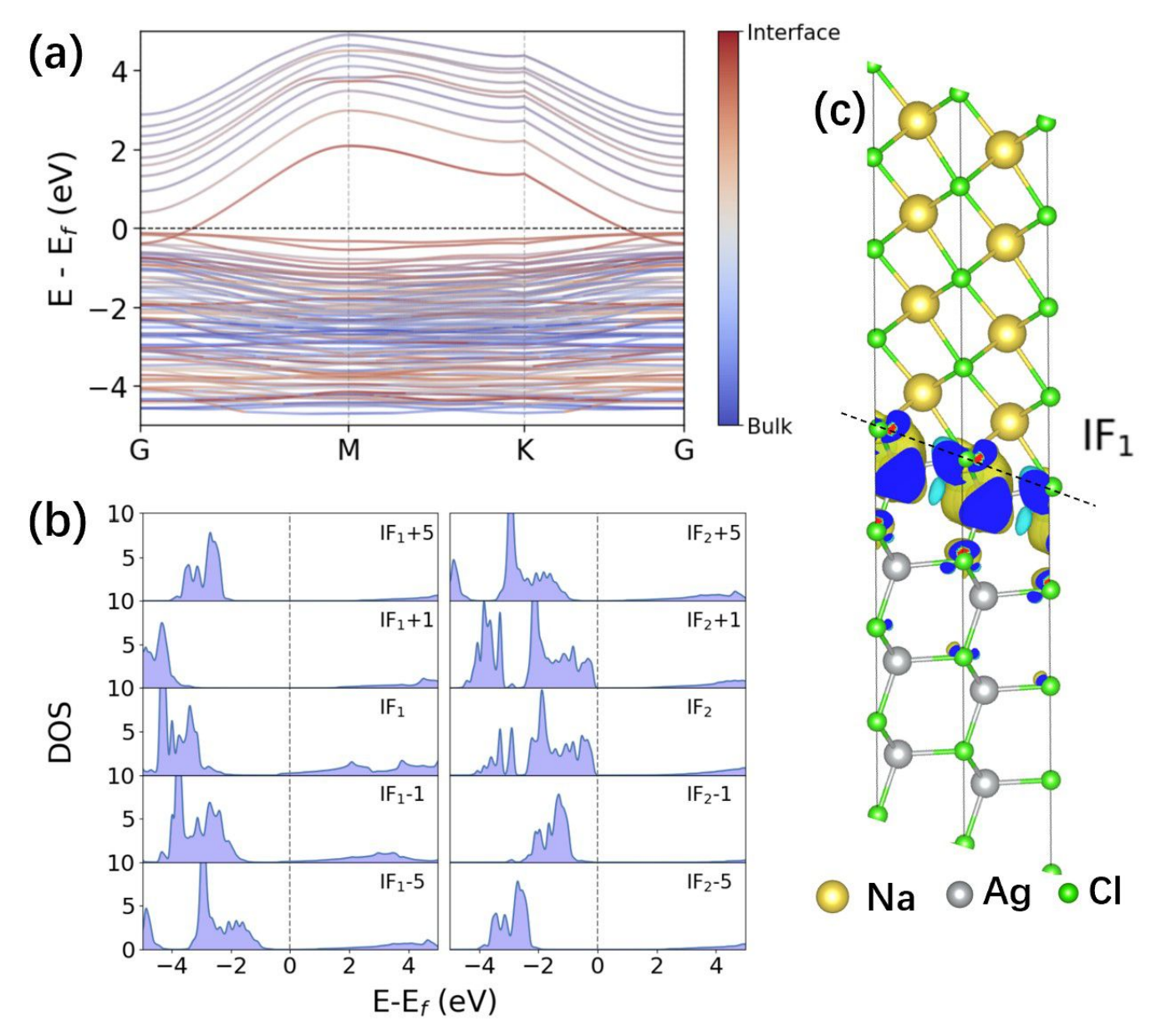}
			\caption{{\bf Electronic structure and interfacial states of the (AgCl)$_{10}$/(NaCl)$_{10}$ superlattice.}
				(a) Band structure of the superlattice, where red denotes states localized at the interfaces and blue corresponds to bulk-like states.
				(b) Layer-resolved PDOS for layers at different distances from IF$_1$ and IF$_2$, with metallic states emerging at IF$_1$.
				(c) Partial charge density of the metallic band near the Fermi level, showing strong electron localization at IF$_1$
				corresponding to a 2DEG.
			}
			\label{AgCl-NaCl-primitive}
		\end{figure}

To explore the physical consequences of QP, we investigate heterostructures composed of materials with distinct QPs using first-principles calculations.
As a representative example, we first examine the interface between two materials with distinct QPs: AgCl and NaCl. As a typical zincblende-type compound, AgCl crystallizes in the F$\bar{4}$3m space group with the T$_d$ point group, whereas NaCl adopts the cubic fcc structure with the Pm$\bar{3}$m space group and the O$_h$ point group.
Conventionally, both structural types are regarded as nonpolar. In particular, zincblende-type materials have long been considered unsuitable for realizing two-dimensional electron gases in semiconductor heterostructures~\cite{morkoc2009}.

Because the value of the QP depends on the choice of unit cell~\cite{pang2025}, we adopt the primitive cell throughout to avoid ambiguity. The results obtained using conventional cells yield the same physical conclusions, which are provided in the Supporting Information (SI).
According to the generalized Neumann’s principle~\cite{pang2025}, the primitive cell of zincblende AgX (X=haloid element) may host 1/4 QP, which is confirmed by first-principles calculations for AgCl.
In contrast, first-principles calculations show that the primitive cell of NaCl, which is inversion symmetric, carries an QP of 1/2~\cite{resta2007}.
Therefore, the two materials possess distinct QPs. To investigate the interface between AgCl and NaCl, we construct an (AgCl)$_{10}$/(NaCl)$_{10}$ superlattice, where the subscript denotes the number of unit cells in each constituent stacked along the $c$ axis. AgCl and NaCl have lattice constants of 5.916 \AA~ and 5.653 \AA, respectively, corresponding to a lattice mismatch of about 4.6\%. The two materials share Cl atoms across the interface.

We perform full structural relaxation of the superlattice and subsequently compute its electronic band structure, as shown in Fig.~\ref{AgCl-NaCl-primitive}(a).
Bulk AgCl and NaCl are insulating, with sizable band gaps of 1.14~eV and 4.91~eV, respectively, as obtained using the PBE functional~\cite{perdew1996} .
Remarkably, when the two materials form an interface, distinct metallic states emerge.

To uncover the origin of the metallic bands, we perform a projected band analysis to resolve the contributions from different atoms. The atoms are classified according to their positions, either near the interface or within the bulk region. As shown by the red solid lines in Fig.~\ref{AgCl-NaCl-primitive}(a), the bands near the Fermi level, particularly those crossing the Fermi surface, originate primarily from atoms located at the interface.

Figure~\ref{AgCl-NaCl-primitive}(b) shows the layer-resolved projected density of states (PDOS) near the two interfaces, IF$_1$ and IF$_2$, whose structures
are shown in Supplementary Fig.~1.
Each interface comprises a AgCl and a NaCl unit cells adjacent to the junction, whereas IF$\pm$1 and IF$\pm$5 correspond to the unit cells one and five lattice spacings away from the interface, respectively.
As shown in Fig.~\ref{AgCl-NaCl-primitive}(b), atoms located exactly at IF$_1$ contribute predominantly to the metallic states at the Fermi level, whereas the PDOS at IF$_1\pm5$ correspond to bulk-like states with a large energy gap.
Similarly, IF$_2$ also hosts localized states near the Fermi energy, while contributions from layers farther from the interfaces rapidly diminish.

Figure~\ref{AgCl-NaCl-primitive}(c) shows the charge density distribution of the metallic band near the Fermi energy at IF$_1$, corresponding to a 2DEG. These states are strongly localized at the interface. The charge distribution slightly leans toward the AgCl side, consistent with the smaller bulk band gap of AgCl, which makes it more prone to forming metallic states.

At the microscopic level, the emergence of metallic interface states originates from the discontinuity of polarization across the boundary.
Such a discontinuity imposes a large electrostatic energy penalty due to the uncompensated Coulomb potential.
To minimize this energy, the system spontaneously develops compensating charges at the interface, which screen the polarization mismatch and stabilize the overall structure.
Alternatively, the electrostatic energy can also be relieved through structural relaxation, where substantial atomic displacements mitigate the polarization discontinuity. In a short-period (AgCl)$_5$/(NaCl)$_5$ superlattice, metallic states appear at the interface before structural relaxation but vanish after relaxation, as a small band gap reopens accompanied by significant lattice distortion across the superlattice and an induced polarization of 0.012 C/m$^2$. Detailed results are provided in Supplementary Fig.~2.

The interfacial phenomena observed in the AgCl/NaCl superlattice exemplify a general principle. Similar behaviors appear across diverse materials, from layered nitrides to zincblende semiconductors, where QP mismatch drives metallic states, bound charges, or lattice reconstructions, establishing a unifying mechanism underlying the rich interfacial physics of a wide range of materials.
More examples are provided in SI.

To further establish the universality of this mechanism, we extend our analysis to more complex oxide systems.
Perovskite oxides, with their structural flexibility and tunable polarization behavior, provide an ideal platform for testing QP-induced interfacial effects beyond simple binary compounds.

Historically, metallic interfaces have been observed in LaAlO$_3$/SrTiO$_3$ heterostructures~\cite{LAO-STO-1,LAO-STO-2,LAO-STO-3}, which are conventionally explained by the polar catastrophe model, where charge transfer compensates for the discontinuity between polar and nonpolar layers.
While this model offers a valuable microscopic explanation, our framework provides a more general, symmetry-based interpretation. It redefines  this classical system as a junction between materials with distinct QPs, unifying the understanding of metallic interface formation beyond specific chemistries or structures.

		\begin{figure}[tbp]
			\centering
			\includegraphics[width=0.45\linewidth]{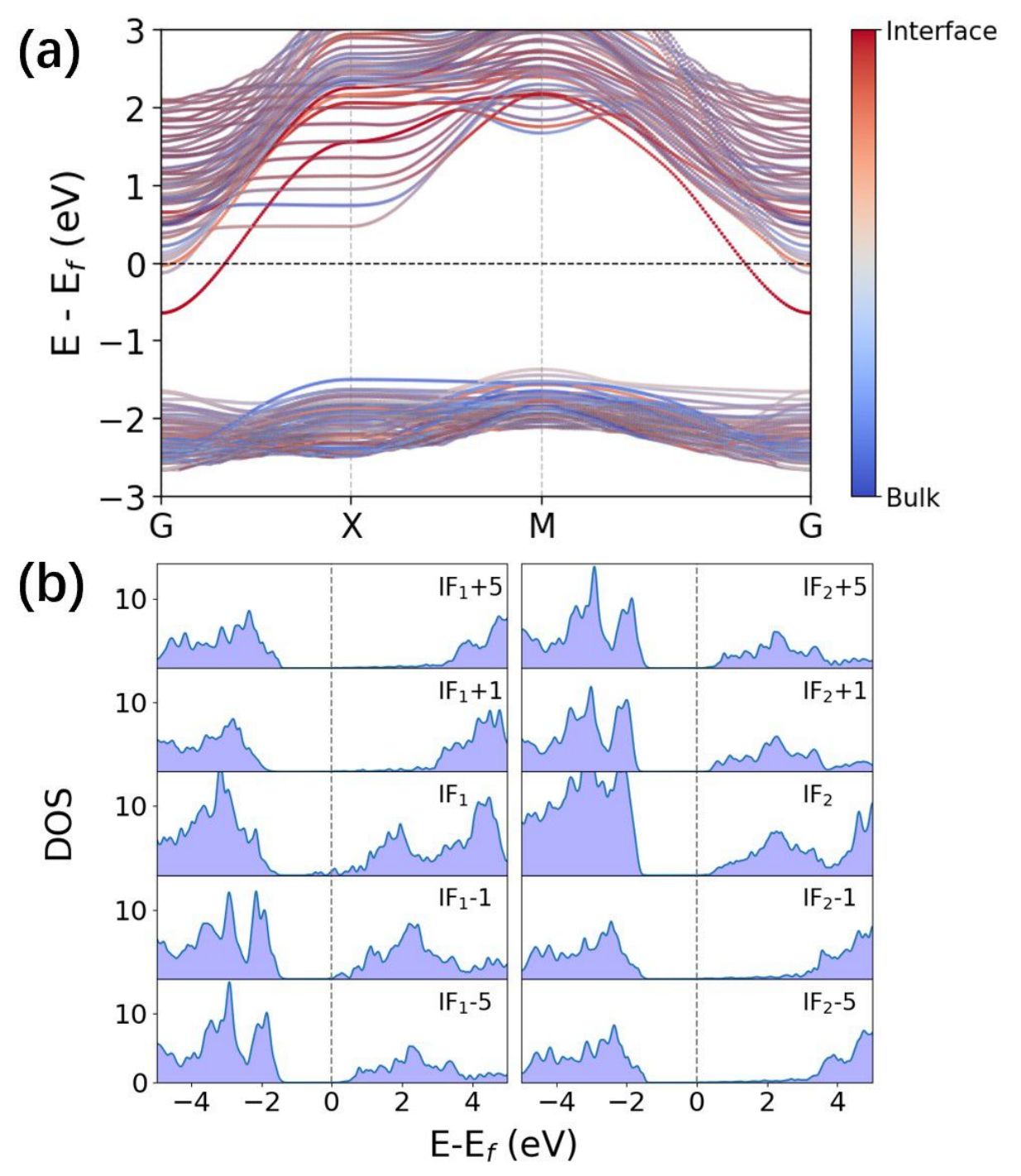}
			\caption{{\bf Electronic structure and interfacial states of the (AgNbO$_3$)$_{10}$/(CaSnO$_3$)$_{10}$ superlattice.}
				(a) Band structure of the superlattice, where red denotes states localized at the interfaces and blue corresponds to bulk-like states.
				(b) Layer-resolved PDOS for atomic layers at varying distances from IF$_1$ and IF$_2$, showing that metallic states emerge at IF$_1$.
			}
			\label{ANO-CSO}
		\end{figure}

We first investigate a perovskite superlattice composed of AgNbO$_3$ and CaSnO$_3$. The lattice parameters of AgNbO$_3$ and CaSnO$_3$ are 3.960 and 3.965~\AA, corresponding to a lattice mismatch of less than 1\%. Both compounds crystallize in the centrosymmetric Pm$\bar{3}$m space group with O$_h$ point symmetry. First-principles calculations yield a 1/2 QP for AgNbO$_3$, while CaSnO$_3$ possesses a zero QP, both prohibiting conventional ferroelectric polarization.
The NbO$_2$-CaO interface is chosen for both IF$_1$ and IF$_2$ to introduce an additional electron, following Ref.~\cite{wang2009first}. The results for other choices of interfacial layers are discussed in the SI.

Figure~\ref{ANO-CSO}(a) shows the band structure, and Fig.~\ref{ANO-CSO}(b) presents the layer-resolved PDOS, revealing a 2DEG localized at the interface of the (AgNbO$_3$)$_{10}$/(CaSnO$_3$)$_{10}$ superlattice. Similar perovskite heterostructures, such as KNbO$_3$/SrTiO$_3$, have been previously reported~\cite{pentcheva2009,yang2016} to host interfacial 2DEGs. However, KNbO$_3$ belongs to the P4mm point group and exhibits conventional ferroelectricity, whereas both AgNbO$_3$ and CaSnO$_3$ are centrosymmetric. This allows us to attribute the emergence of the 2DEG purely to the QP mismatch, rather than to traditional ferroelectric polarization.
		
		\begin{figure}[tbp]
			\centering
			\includegraphics[width=0.45\linewidth]{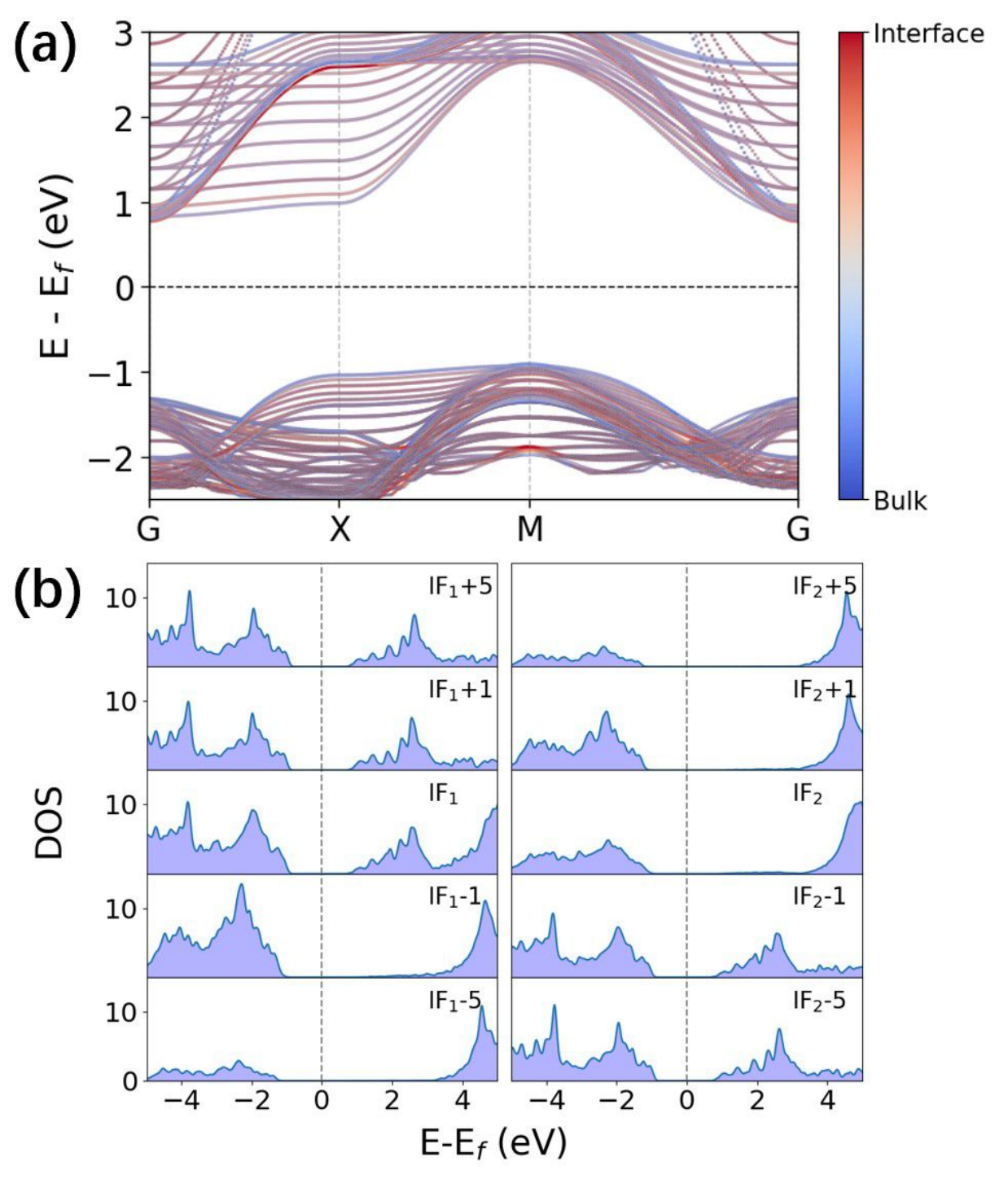}
			\caption{{\bf Electronic structure of the (SrTiO$_3$)$_{10}$/(CaSnO$_3$)$_{10}$ superlattice.}
				(a) Band structure of the superlattice revealing a large band gap with no interface states appearing near the Fermi level.
				(b) Layer-resolved PDOS at varying distances from IF$_1$ and IF$_2$, indicating close similarity to the bulk electronic structure
				with large band gap.
			}
						\label{CSO-STO}
		\end{figure}
			
By contrast, both CaSnO$_3$ and SrTiO$_3$ possess zero QP, therefore, no polarization discontinuity is expected, and consequently, no pronounced interfacial effects occur. The calculations show that both types of interfaces, TiO$_2$–CaO and SnO$_2$–SrO, yield essentially the same results, with no localized states appearing near the Fermi level.
The band structure and layer-resolved PDOS, shown in Fig.~\ref{CSO-STO}(a) and (b), confirm that the heterostructure remains insulating, with a band gap of 2.06 eV, comparable to that of bulk SrTiO$_3$ (2.00 eV). This further supports that the discontinuity in QP is the driving factor for 2DEG formation. A similar case is observed in the LaAlO$_3$/BiGaO$_3$ heterostructure, where both materials possess 1/2 QP. Owing to the absence of polarization discontinuity, no metallic states emerge, and the bands are dominated by bulk-like states, as shown in the Supplementary Fig.~7.

However, because polarization is defined only up to an integer multiple of the polarization quantum, the situation becomes more subtle. For instance, consider a superlattice composed of LaAlO$_3$ and AgNbO$_3$. Both materials nominally possess a QP of 1/2. Yet, using a simple point-charge model with $Z_{\mathrm{La}} = +3$, $Z_{\mathrm{Al}} = +3$, $Z_{\mathrm{O}} = -2$, and $Z_{\mathrm{Ag}} = +1$, $Z_{\mathrm{Nb}} = +5$, we obtain QPs of 1/2 and –1/2 for LaAlO$_3$ and AgNbO$_3$, respectively. Their QPs thus differ by one quantum, producing a net polarization discontinuity at the interface that can drive the formation of a 2DEG, as shown in Supplementary Fig.~12.


To conclude, this work establishes the QP as a genuine, symmetry-protected invariant of crystalline solids, rather than a purely formal construct.
We demonstrate that discontinuities in QP between distinct materials can induce a wide spectrum of phenomena, including two-dimensional metallic states, bound charges, and structural reconstructions.
In several systems, the polarization discontinuity is sufficiently strong that the resulting electronic and lattice responses extend deep into the bulk, giving rise to long-range distortion and even metallic states away from the interface.
These findings redefine the fundamental understanding of polar interfaces and introduce a unified framework for symmetry-driven interfacial behavior.
Beyond their conceptual significance, they also open new pathways for the rational design of functional quantum materials through controlled quantized polarization mismatch.

\section*{Methods}
\label{sec:method}
The first-principles calculations are performed within the framework of density functional theory (DFT) using the Atomic Orbital-Based Ab-initio Computation at USTC (ABACUS) package \cite{chen2010,li2016}. The ABACUS code is optimized for large-scale DFT simulations and employs numerical atomic orbital (NAO) basis sets \cite{linpz2021,Linpz2023}. In this study, the Perdew–Burke–Ernzerhof (PBE) \cite{perdew1996} functional is used to describe exchange–correlation effects. Optimized norm-conserving Vanderbilt (ONCV) pseudopotentials from the SG15 library \cite{Scherpelz2016SG15} are adopted. Details of the valence configurations and basis orbitals for each element are provided in the SI.
During both structural relaxation and self-consistent calculations, an $8\times8\times1$ Monkhorst–Pack $k$-point mesh is employed, with atomic layers stacked along the $c$-axis. The plane-wave energy cutoff for the wave functions is set to 100 Ry.
All heterostructures are fully relaxed until the residual pressure is less than 0.5 kbar and the maximum force on any atom is less than 0.05 eV/\AA~,
unless stated otherwise.	
The electric polarization is computed using the modern theory of polarization~\cite{KingSmith1993}, as implemented in the PYATB package~\cite{jin2023}.
PYATB is also employed for the calculation of band structures and PDOS, utilizing the tight-binding Hamiltonian constructed directly from the self-consistent outputs of ABACUS. A $30\times30\times1$ $k$-point mesh is adopted for PDOS calculations for the superlattices.

\section*{Data availability}

All data generated and/or analysed during this study are included in this article.

\section*{Code availability}

The ABACUS code is an open source DFT code under the GPL 3.0 licence, which is available from http://abacus.ustc.edu.cn.
The Py-ATB code, also under the GPL 3.0 licence, can be downloaded from https://github.com/pyatb.

\section*{Acknowledgements}
This work was supported by the Strategic Priority Research Program of Chinese Academy of Sciences (Grant Number XDB0500201), the National Natural Science Foundation of China (Grant Number 12134012),
and the Innovation Program for Quantum Science and Technology Grant Number 2021ZD0301200.
The numerical calculations were performed on the USTC HPC facilities.

\section*{Author contributions}

L. He conceived the idea and led the project. H. Pang performed the calculations under the supervision of L. He. Both authors analyzed the results and contributed to writing the manuscript.

\section*{Competing interests}

The authors declare no competing interests.


\begin{thebibliography}{10}
\expandafter\ifx\csname url\endcsname\relax
  \def\url#1{\texttt{#1}}\fi
\expandafter\ifx\csname urlprefix\endcsname\relax\def\urlprefix{URL }\fi
\providecommand{\bibinfo}[2]{#2}
\providecommand{\eprint}[2][]{\url{#2}}

\bibitem{cochran1960}
\bibinfo{author}{Cochran, W.}
\newblock \bibinfo{title}{Crystal stability and the theory of
  ferroelectricity}.
\newblock \emph{\bibinfo{journal}{Adv. Phys.}} \textbf{\bibinfo{volume}{9}},
  \bibinfo{pages}{387} (\bibinfo{year}{1960}).

\bibitem{portengen1996}
\bibinfo{author}{Portengen, T.}, \bibinfo{author}{{\"O}streich, T.} \&
  \bibinfo{author}{Sham, L.}
\newblock \bibinfo{title}{Theory of electronic ferroelectricity}.
\newblock \emph{\bibinfo{journal}{Phys. Rev. B}} \textbf{\bibinfo{volume}{54}},
  \bibinfo{pages}{17452} (\bibinfo{year}{1996}).

\bibitem{lines2001}
\bibinfo{author}{Lines, M.~E.} \& \bibinfo{author}{Glass, A.~M.}
\newblock \emph{\bibinfo{title}{Principles and applications of ferroelectrics
  and related materials}} (\bibinfo{publisher}{Oxford University Press},
  \bibinfo{address}{Oxford, UK}, \bibinfo{year}{2001}).

\bibitem{rabe2007physics}
\bibinfo{author}{Rabe, K.~M.}, \bibinfo{author}{Ahn, C.~H.} \&
  \bibinfo{author}{Triscone, J.-M.}
\newblock \emph{\bibinfo{title}{Physics of ferroelectrics: a modern
  perspective}}, vol. \bibinfo{volume}{105} (\bibinfo{publisher}{Springer
  Science \& Business Media}, \bibinfo{year}{2007}).

\bibitem{neumann1885vorlesungen}
\bibinfo{author}{Neumann, F.}
\newblock \emph{\bibinfo{title}{Vorlesungen {\"u}ber die Theorie der
  Elasticit{\"a}t der festen K{\"o}rper und des Licht{\"a}thers}},
  vol.~\bibinfo{volume}{5} (\bibinfo{publisher}{BG Teubner},
  \bibinfo{year}{1885}).

\bibitem{Ji2024}
\bibinfo{author}{Ji, J.}, \bibinfo{author}{Yu, G.}, \bibinfo{author}{Xu, C.} \&
  \bibinfo{author}{Xiang, H.}
\newblock \bibinfo{title}{Fractional quantum ferroelectricity}.
\newblock \emph{\bibinfo{journal}{Nat. Commun.}} \textbf{\bibinfo{volume}{15}},
  \bibinfo{pages}{135} (\bibinfo{year}{2024}).

\bibitem{Yu2025}
\bibinfo{author}{Yu, G.}, \bibinfo{author}{Ji, J.}, \bibinfo{author}{Chen, Y.},
  \bibinfo{author}{Xu, C.} \& \bibinfo{author}{Xiang, H.}
\newblock \bibinfo{title}{Symmetry strategy for rapid discovery of abundant
  fractional quantum ferroelectrics}.
\newblock \emph{\bibinfo{journal}{Phys. Rev. Lett.}}
  \textbf{\bibinfo{volume}{134}}, \bibinfo{pages}{016801}
  (\bibinfo{year}{2025}).

\bibitem{pang2025}
\bibinfo{author}{Pang, H.} \& \bibinfo{author}{He, L.}
\newblock \bibinfo{title}{Generalized neumann’s principle as a unified
  framework for fractional quantum and conventional ferroelectricity}.
\newblock \emph{\bibinfo{journal}{Phys. Rev. Lett.}}
  \textbf{\bibinfo{volume}{135}}, \bibinfo{pages}{116402}
  (\bibinfo{year}{2025}).

\bibitem{vanderbilt1993electric}
\bibinfo{author}{Vanderbilt, D.} \& \bibinfo{author}{King-Smith, R.}
\newblock \bibinfo{title}{Electric polarization as a bulk quantity and its
  relation to surface charge}.
\newblock \emph{\bibinfo{journal}{Phys. Rev. B}} \textbf{\bibinfo{volume}{48}},
  \bibinfo{pages}{4442} (\bibinfo{year}{1993}).

\bibitem{vanderbilt2018berry}
\bibinfo{author}{Vanderbilt, D.}
\newblock \emph{\bibinfo{title}{Berry phases in electronic structure theory:
  electric polarization, orbital magnetization and topological insulators}}
  (\bibinfo{publisher}{Cambridge University Press},
  \bibinfo{address}{Cambridge, UK}, \bibinfo{year}{2018}).

\bibitem{chern}
\bibinfo{author}{Hatsugai, Y.}
\newblock \bibinfo{title}{Chern number and edge states in the integer quantum
  hall effect}.
\newblock \emph{\bibinfo{journal}{Phys. Rev. Lett.}}
  \textbf{\bibinfo{volume}{71}}, \bibinfo{pages}{3697} (\bibinfo{year}{1993}).

\bibitem{xiao2010berry}
\bibinfo{author}{Xiao, D.}, \bibinfo{author}{Chang, M.-C.} \&
  \bibinfo{author}{Niu, Q.}
\newblock \bibinfo{title}{Berry phase effects on electronic properties}.
\newblock \emph{\bibinfo{journal}{Rev. Mod. Phys.}}
  \textbf{\bibinfo{volume}{82}}, \bibinfo{pages}{1959--2007}
  (\bibinfo{year}{2010}).

\bibitem{LAO-STO-1}
\bibinfo{author}{Nakagawa, N.}, \bibinfo{author}{Hwang, H.~Y.} \&
  \bibinfo{author}{Muller, D.~A.}
\newblock \bibinfo{title}{Why some interfaces cannot be sharp}.
\newblock \emph{\bibinfo{journal}{Nat. Mater.}} \textbf{\bibinfo{volume}{5}},
  \bibinfo{pages}{204--209} (\bibinfo{year}{2006}).

\bibitem{LAO-STO-2}
\bibinfo{author}{Zhang, F.} \emph{et~al.}
\newblock \bibinfo{title}{Modulating the electrical transport in the
  two-dimensional electron gas at {LaAlO$_3$/SrTiO$_3$} heterostructures by
  interfacial flexoelectricity}.
\newblock \emph{\bibinfo{journal}{Phys. Rev. Lett.}}
  \textbf{\bibinfo{volume}{122}}, \bibinfo{pages}{257601}
  (\bibinfo{year}{2019}).

\bibitem{LAO-STO-3}
\bibinfo{author}{Ohtomo, A.} \& \bibinfo{author}{Hwang, H.}
\newblock \bibinfo{title}{A high-mobility electron gas at the
  {LaAlO$_3$}/{SrTiO$_3$} heterointerface}.
\newblock \emph{\bibinfo{journal}{Nature}} \textbf{\bibinfo{volume}{427}},
  \bibinfo{pages}{423} (\bibinfo{year}{2004}).

\bibitem{morkoc2009}
\bibinfo{author}{Morko{\c{c}}, H.}
\newblock \emph{\bibinfo{title}{Handbook of nitride semiconductors and devices,
  Materials Properties, Physics and Growth}} (\bibinfo{publisher}{John Wiley \&
  Sons}, \bibinfo{year}{2009}).

\bibitem{resta2007}
\bibinfo{author}{Resta, R.} \& \bibinfo{author}{Vanderbilt, D.}
\newblock \bibinfo{title}{Theory of polarization: a modern approach}.
\newblock In \emph{\bibinfo{booktitle}{Physics of ferroelectrics: a modern
  perspective}}, \bibinfo{pages}{31--68} (\bibinfo{publisher}{Springer},
  \bibinfo{year}{2007}).

\bibitem{perdew1996}
\bibinfo{author}{Perdew, J.~P.}, \bibinfo{author}{Burke, K.} \&
  \bibinfo{author}{Ernzerhof, M.}
\newblock \bibinfo{title}{Generalized gradient approximation made simple}.
\newblock \emph{\bibinfo{journal}{Phys. Rev. Lett.}}
  \textbf{\bibinfo{volume}{77}}, \bibinfo{pages}{3865--3868}
  (\bibinfo{year}{1996}).

\bibitem{wang2009first}
\bibinfo{author}{Wang, Y.}, \bibinfo{author}{Niranjan, M.~K.},
  \bibinfo{author}{Jaswal, S.} \& \bibinfo{author}{Tsymbal, E.~Y.}
\newblock \bibinfo{title}{First-principles studies of a two-dimensional
  electron gas at the interface in ferroelectric oxide heterostructures}.
\newblock \emph{\bibinfo{journal}{Phys. Rev. B: Condens. Matter}}
  \textbf{\bibinfo{volume}{80}}, \bibinfo{pages}{165130}
  (\bibinfo{year}{2009}).

\bibitem{pentcheva2009}
\bibinfo{author}{Pentcheva, R.} \& \bibinfo{author}{Pickett, W.~E.}
\newblock \bibinfo{title}{Avoiding the polarization catastrophe in {LaAlO$_3$}
  in overlayers on {SrTiO$_3$} (001) through polar distortion}.
\newblock \emph{\bibinfo{journal}{Phys. Rev. Lett.}}
  \textbf{\bibinfo{volume}{102}}, \bibinfo{pages}{107602}
  (\bibinfo{year}{2009}).

\bibitem{yang2016}
\bibinfo{author}{Yang, K.}, \bibinfo{author}{Nazir, S.},
  \bibinfo{author}{Behtash, M.} \& \bibinfo{author}{Cheng, J.}
\newblock \bibinfo{title}{High-throughput design of two-dimensional electron
  gas systems based on polar/nonpolar perovskite oxide heterostructures}.
\newblock \emph{\bibinfo{journal}{Sci. Rep.}} \textbf{\bibinfo{volume}{6}},
  \bibinfo{pages}{34667} (\bibinfo{year}{2016}).

\bibitem{chen2010}
\bibinfo{author}{Chen, M.}, \bibinfo{author}{Guo, G.-C.} \&
  \bibinfo{author}{He, L.}
\newblock \bibinfo{title}{Systematically improvable optimized atomic basis sets
  for ab initio calculations}.
\newblock \emph{\bibinfo{journal}{J. Phys.: Condens. Matter}}
  \textbf{\bibinfo{volume}{22}}, \bibinfo{pages}{445501}
  (\bibinfo{year}{2010}).

\bibitem{li2016}
\bibinfo{author}{Li, P.} \emph{et~al.}
\newblock \bibinfo{title}{Large-scale ab initio simulations based on
  systematically improvable atomic basis}.
\newblock \emph{\bibinfo{journal}{Comput. Mater. Sci.}}
  \textbf{\bibinfo{volume}{112}}, \bibinfo{pages}{503--517}
  (\bibinfo{year}{2016}).

\bibitem{linpz2021}
\bibinfo{author}{Lin, P.}, \bibinfo{author}{Ren, X.} \& \bibinfo{author}{He,
  L.}
\newblock \bibinfo{title}{Strategy for constructing compact numerical atomic
  orbital basis sets by incorporating the gradients of reference
  wavefunctions}.
\newblock \emph{\bibinfo{journal}{Phys. Rev. B}}
  \textbf{\bibinfo{volume}{103}}, \bibinfo{pages}{235131}
  (\bibinfo{year}{2021}).

\bibitem{Linpz2023}
\bibinfo{author}{Lin, P.}, \bibinfo{author}{Ren, X.}, \bibinfo{author}{Liu, X.}
  \& \bibinfo{author}{He, L.}
\newblock \bibinfo{title}{Ab initio electronic structure calculations based on
  numerical atomic orbitals: Basic fomalisms and recent progresses}.
\newblock \emph{\bibinfo{journal}{WIREs Comput. Mol. Sci.}}
  \textbf{\bibinfo{volume}{14}}, \bibinfo{pages}{e1687} (\bibinfo{year}{2023}).

\bibitem{Scherpelz2016SG15}
\bibinfo{author}{Scherpelz, P.}, \bibinfo{author}{Govoni, M.},
  \bibinfo{author}{Hamada, I.} \& \bibinfo{author}{Galli, G.}
\newblock \bibinfo{title}{Implementation and validation of fully relativistic
  gw calculations: Spin–orbit coupling in molecules, nanocrystals, and
  solids}.
\newblock \emph{\bibinfo{journal}{J. Chem. Theory Comput.}}
  \textbf{\bibinfo{volume}{12}}, \bibinfo{pages}{3523--3544}
  (\bibinfo{year}{2016}).

\bibitem{KingSmith1993}
\bibinfo{author}{King-Smith, R.~D.} \& \bibinfo{author}{Vanderbilt, D.}
\newblock \bibinfo{title}{Theory of polarization of crystalline solids}.
\newblock \emph{\bibinfo{journal}{Phys. Rev. B}} \textbf{\bibinfo{volume}{47}},
  \bibinfo{pages}{1651--1654} (\bibinfo{year}{1993}).

\bibitem{jin2023}
\bibinfo{author}{Jin, G.}, \bibinfo{author}{Pang, H.}, \bibinfo{author}{Ji,
  Y.}, \bibinfo{author}{Dai, Z.} \& \bibinfo{author}{He, L.}
\newblock \bibinfo{title}{Pyatb: An efficient python package for electronic
  structure calculations using ab initio tight-binding model}.
\newblock \emph{\bibinfo{journal}{Comput. Phys. Commun.}}
  \textbf{\bibinfo{volume}{291}}, \bibinfo{pages}{108844}
  (\bibinfo{year}{2023}).

\end{thebibliography}

\end{document}